\documentstyle[11pt,newpasp,twoside,epsf]{article}
\newcommand{\hi}{H{\sc i}}
\newcommand{\lya}{Lyman $\alpha$}

\newcommand{\rjup}{R$_{\rm Jup}$}
\newcommand{\tup}{T$_{\rm up}$}
\newcommand{\teff}{T$_{\rm eff}$}

\def\edcomment#1{\iffalse\marginpar{\raggedright\sl#1\/}\else\relax\fi}
\reversemarginpar

\begin{document}

\title{``Osiris''(HD209458b), an evaporating planet}

\author{Alfred  Vidal-Madjar \& Alain Lecavelier des Etangs}
\affil{Institut d'Astrophysique de Paris du CNRS, 98bis Boulevard Arago,
75014, Paris, FRANCE}

\begin{abstract}
Three transits of the planet orbiting the solar type star
HD\,209458 were observed in the far UV at the wavelength of the
\hi\ \lya\ line. The planet size at this wavelength is equal to
4.3~\rjup , i.e. larger than the planet Roche radius (3.6~\rjup ).
Absorbing hydrogen atoms were found to be blueshifted by up to
--130~km.s$^{-1}$, exceeding the planet escape velocity. This
implies that hydrogen atoms are escaping this ``hot Jupiter'' planet.
An escape flux of $\ga$10$^{10}$~g.s$^{-1}$ is needed to explain 
the observations. Taking into account the tidal forces and 
the temperature rise expected in the upper atmosphere, 
theoretical evaluations are in good agreement with the observed rate.
Lifetime of planets closer to their star could be
shorter than stellar lifetimes suggesting that this evaporating
phenomenon may explain the lack of planets with very short orbital
distance. 

This evaporating planet could be represented by the Egyptian God
``Osiris'' cut into pieces and having lost one of them. This would
give us a much easier way to name that planet and replace the
unpleasant ``HD209458b'' name used so far.

\end{abstract}

\section{Introduction}

The extrasolar planet in the system HD209458 is the first one for
which repeated transits across the stellar disk have been observed
($\sim$1.35\%\ absorption; Henry et al., 2000; Charbonneau et al.,
2000). Together with radial velocity measurements (Mazeh et al.,
2000), this has led to a determination of the planet's radius and
mass, confirming that it must be a gas giant. The transits of this
gaseous extrasolar planet offer a unique
opportunity to investigate the spectral features of its
atmosphere. Numerous searches for an atmospheric signature in the
IR-optical wavelength range failed until the detection with Hubble
Space Telescope (HST) of the \underline{dense lower} atmosphere of
HD209458b observed in the neutral sodium lines
($\sim$0.02\%\ additional absorption; Charbonneau et al., 2002).

Far more abundant than any other species, hydrogen is well suited
for searching atmospheric absorptions during transits, in particular 
over the stellar ultraviolet (UV) \lya\ emission line at 1215.67\AA.
With recent HST observations in the UV, Vidal--Madjar et al.
(2003) detected the H\,{\sc i} atomic hydrogen absorption
($\sim$15\% ) over the stellar \lya\ line during three transits of
HD209458b. Comparison with models showed that this absorption
should take place beyond the Roche limit and could thus be
understood only in terms of escaping hydrogen atoms.

The detection of this \underline{extended upper} atmosphere (the
exosphere) provides a new and crucial step in our understanding of the
outer atmospheric structures of these ``hot'' extrasolar planets.

\begin{figure}
\plotfiddle{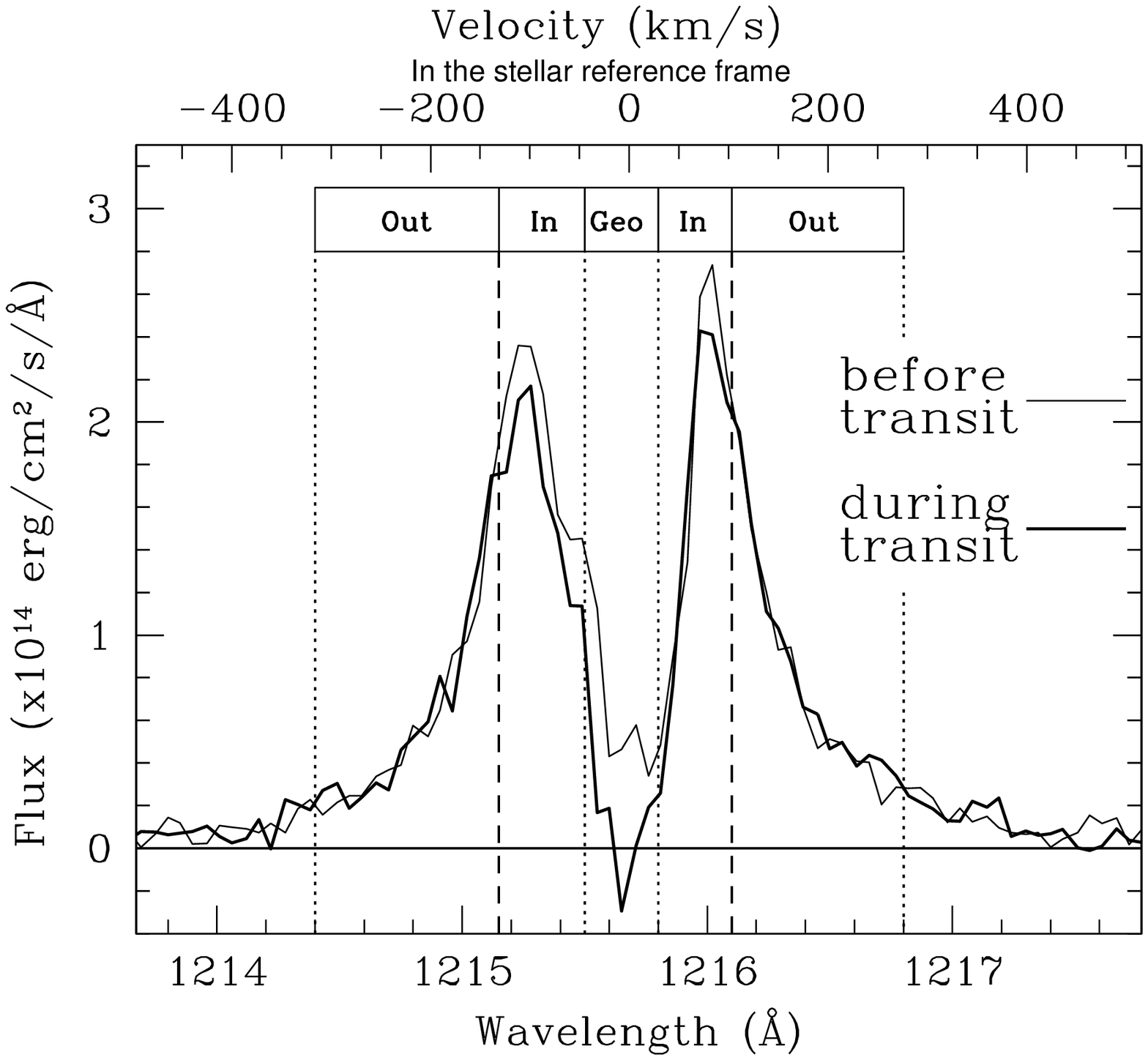}{0cm}{0}{30}{30}{-175}{-195}
\plotfiddle{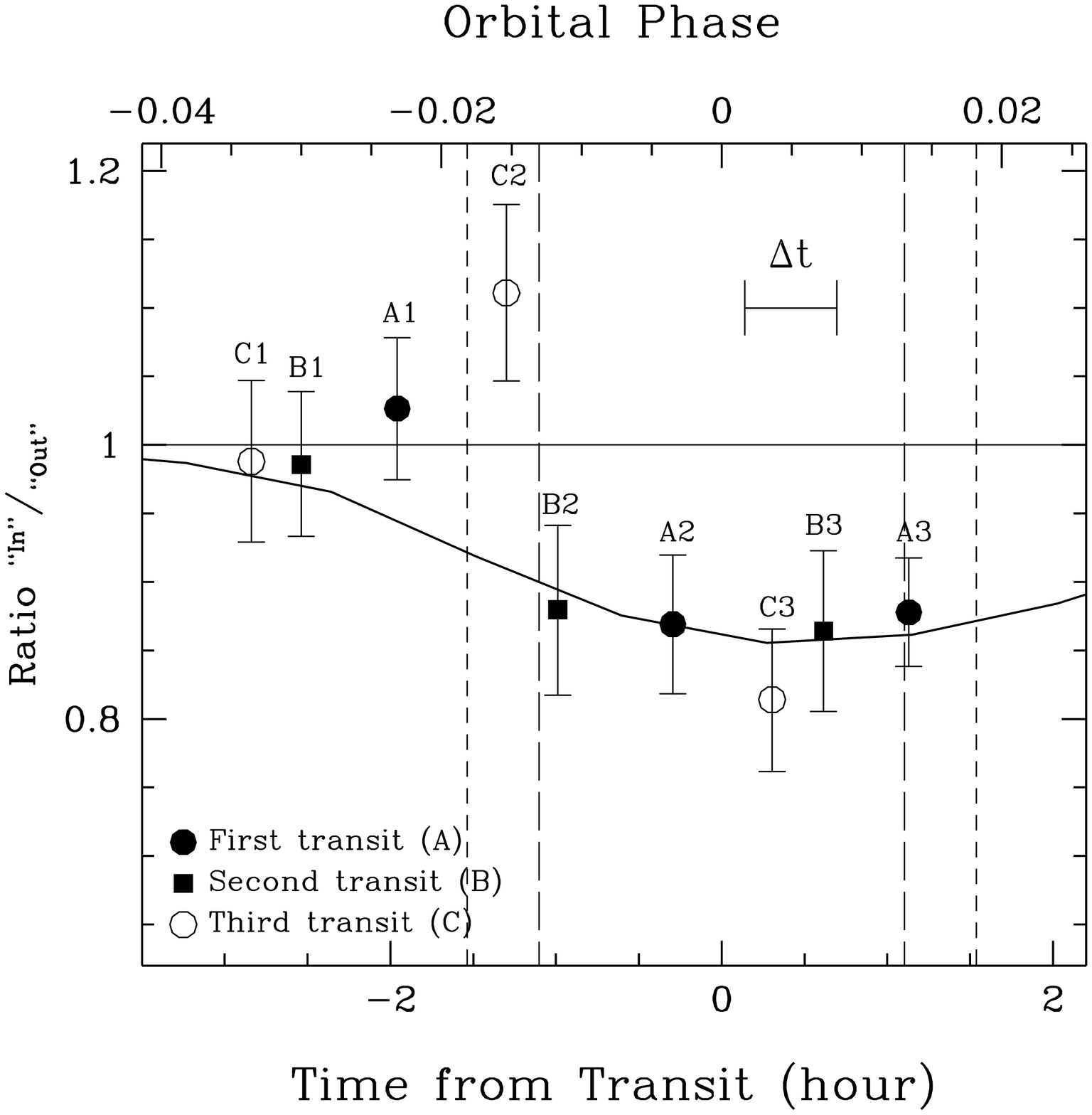}{3.5cm}{0}{25.5}{25.5}{20}{-49}
\caption{{\bf
Left:} The \lya\ stellar line as observed by Vidal--Madjar et al.
(2003). The averaged profile observed during transit (thick line)
presents a reduced flux when compared to the pre--transit profile
(thin line).
The region named ``Geo'' corresponds to
the region where the geocoronal \lya\ correction was too important.
In the ``In'' region absorption is observed while the ``Out'' region serves
as a flux reference.
{\bf Right:} The averaged ``In''/``Out'' flux ratio in the
individual exposures of the three observed transits
(see text).
Exposures A1, B1, and C1 were performed
before  and A2, B3, and C3 entirely during transits.
Error bars are $\pm1\sigma$. The ``In''/``Out'' ratio 
decreases by $\sim$15\% during the transit. The thick line
represents the absorption ratio modeled through a particle
simulation (see Fig.~3).}
\end{figure}

\section{The observations}

Three transits of HD209458b (noted A, B and C) 
were surveyed with the STIS
spectrograph on-board HST ($\sim$20~km.s$^{-1}$ resolution). 
For each transit, three consecutive HST
orbits were scheduled such that the first orbit ended before the
first contact to serve as a reference, the two following ones
being partly or entirely within the transit.
An average 15$\pm$4\% (1$\sigma$) relative intensity drop
(variation of the ratio between the flux in the ``In'' over the
flux in the ``Out'' region, noted ``In''/``Out'', see Fig.~1) near
the center of the \lya\ line was observed during the transits.
This is larger than expected for the atmosphere of a planet
occulting only $\sim$1.5\% of the star. Note that this drop has to
be added on top of the $\sim$1.5\% reduction of the whole stellar
spectrum caused by the planetary disk transit which includes of
course the \lya\ line.

Because of the small distance (8.5~R$_*$) between the planet and the star
(allowing an intense heating of the planet and its classification
as a ``hot Jupiter'') the Roche lobe is only at 2.7 planetary radii 
(i.e. 3.6~\rjup ). Filling up this lobe with hydrogen atoms gives a
maximum absorption of $\sim$10\% during planetary transits. Since
a more important absorption was detected, hydrogen atoms cover a
larger area corresponding to a spherical object of 4.3~\rjup .
Observed beyond the Roche limit, these hydrogen atoms must be escaping 
the planet. Independently, the spectral absorption width (between
$-130$ and $+100$ km.s$^{-1}$, see Fig.~1) also shows that some
hydrogen atoms have large velocities relative to the planet, exceeding
the escape velocity. This further confirms that hydrogen atoms 
must be escaping the planetary atmosphere.

\begin{figure}
\plotfiddle{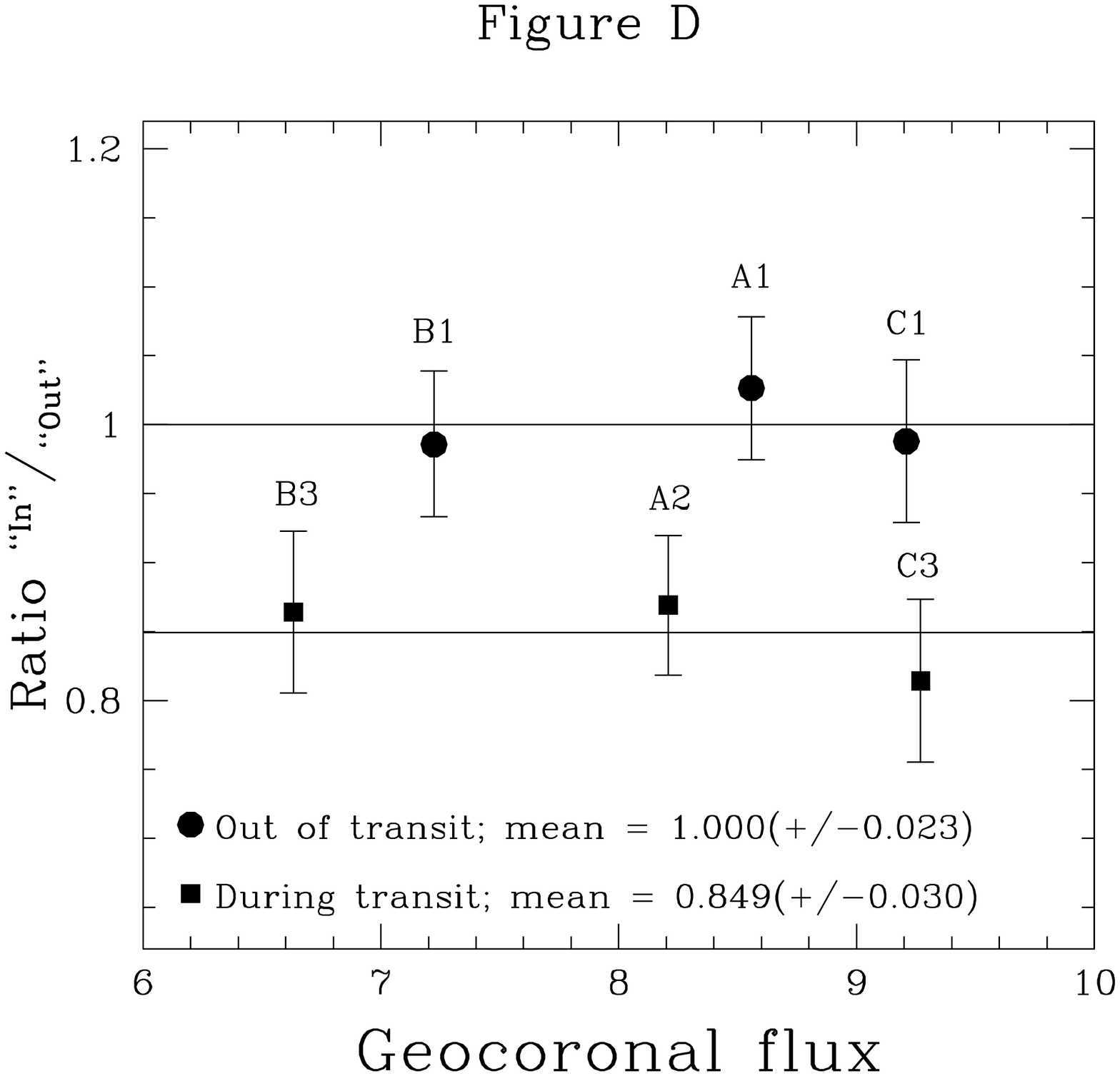}{0cm}{0}{27}{27}{-170}{-165}
\plotfiddle{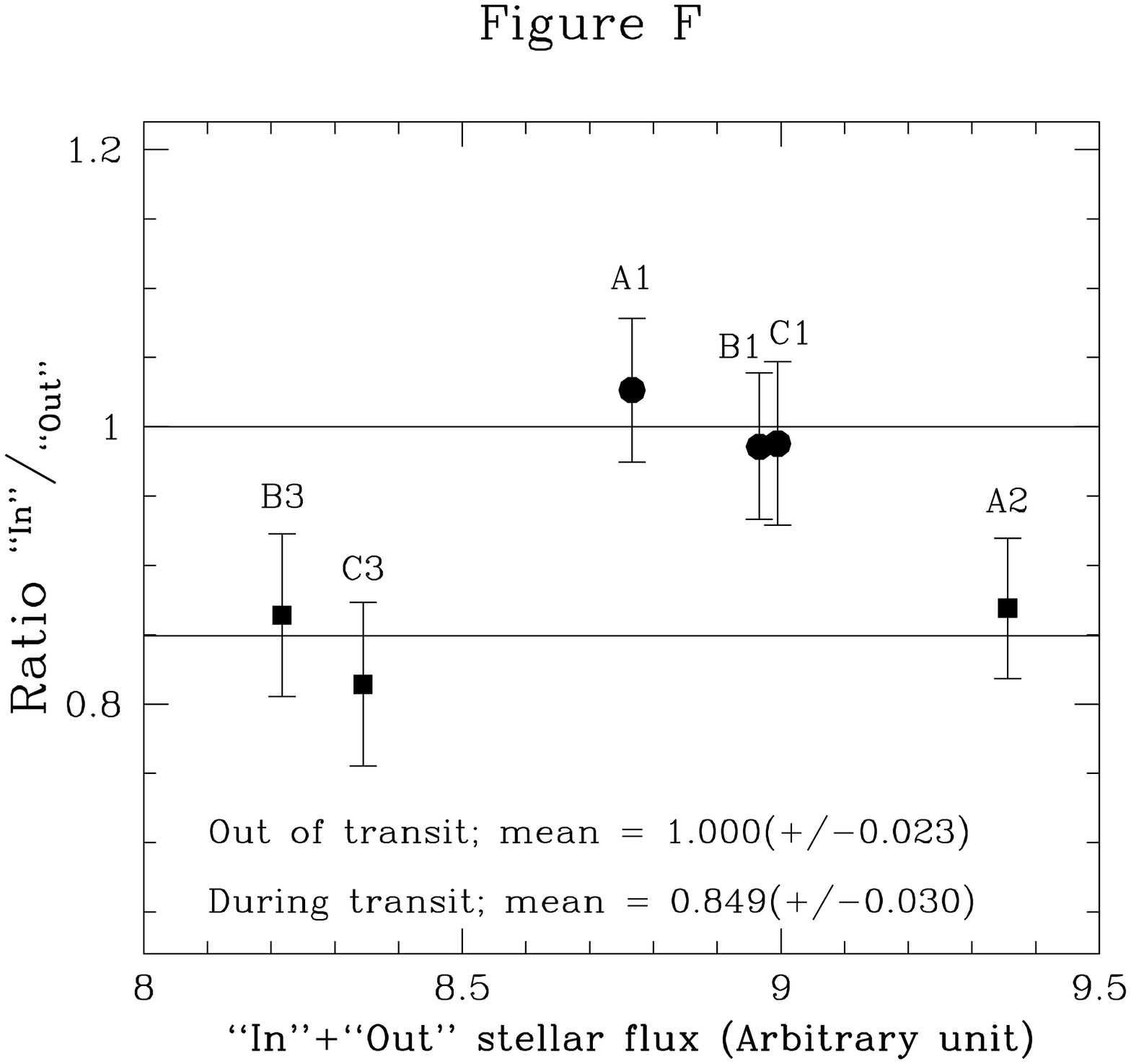}{3.5cm}{0}{27}{27}{-10}{-53}
\caption{The ``In''/``Out'' ratios corresponding to the
A1, B1 and C1 observations performed out
of the transits and to the A2, B3 and C3 ones completely
within the transits are shown (see text and Fig.~1). 
{\bf Left:} In Figure D the ratios are shown 
as a function of the corresponding geocoronal \lya\ emission.
{\bf Right:} Figure F shows the same ratios as a
function of the total stellar \lya\ flux. In both cases the
absorption observed during transits is not correlated with
the potential perturbations revealing that they were either properly
corrected (see D\'esert et al. 2004) or irrelevant.}
\end{figure}

\subsection{Observational difficulties}

These are essentially twofold~: flux variations during transits
could be either due to changes in the geocoronal \lya\ emission
which covers and perturbs part of the stellar line or to intrinsic
stellar flux variations.

First the geocoronal \lya\ emission produced by the hydrogen atoms
present above HST in the Earth upper atmosphere (see e.g.
Vidal--Madjar \& Thomas, 1978) fills the aperture of the
spectrograph. This results in an extended emission spread over the
instrument detector, perpendicular to the dispersion direction.
The extent of this emission along the slit allows however a
precise correction at the position of the stellar spectrum. It was
demonstrated that the geocoronal correction was made with high
enough accuracy at least outside the central region ($-40$ and
$+35$ km.s$^{-1}$, noted ``Geo'' in Fig.~1). No correlation was
found between the geocoronal variations and the stellar \lya\
profile changes observed during transits showing independently
that the corrections were properly made (see Fig.~2).

Second the observed \lya\ spectrum of HD209458 (G0V) is typical
for a solar type star, with a double peaked emission originating
in the stellar chromosphere. Stellar variability is thus a concern
because it is well known that solar type stars present important
flux variations in the far and extreme UV as it was observed in
details in the case of the Sun (see e.g. Vidal--Madjar 1975).
The total solar
\lya\ flux varies by about a factor two, while its relative
fluctuations in similar spectral domains (variations of the
``In''/``Out'' ratios) are smaller than $\pm$6\%. Within a few
months, a time comparable to the HD209458 observations, the
``In''/``Out'' solar flux ratio varies by less than $\pm$4\% . If
both stars behave in similar manners, this is an indication that
the absorption detected is not of stellar origin. This is
confirmed by the fact that no correlation was found between the
total stellar flux changes (``In''+``Out'' flux) and the
absorption observed during transits (Fig.~2) showing that the
observed variations are not related to stellar flux fluctuations.

\section{The upper atmospheric model}

The observed 15\% intensity drop could only be explained if
hydrogen atoms are able to reach the Roche lobe of the planet and
then escape. To evaluate the amount of escaping atoms
Vidal--Madjar et al. (2003) have built a particle simulation in
which they assumed that hydrogen atoms are sensitive to the
stellar gravity and radiation pressure (see Fig.~3). In their
simulation, escaping hydrogen atoms expand in an asymmetric
cometary like tail and are progressively ionized when moving away
from the planet. Atoms in the evaporating coma and tail cover a
large area of the star. A minimum escape flux of
$\sim~10^{10}$~g.s$^{-1}$ is compatible with the observations.
However, due to saturation effects in the \lya\ absorption line,
larger escape fluxes by several orders of magnitude would still be
compatible with the observations.

\begin{figure}
\plotfiddle{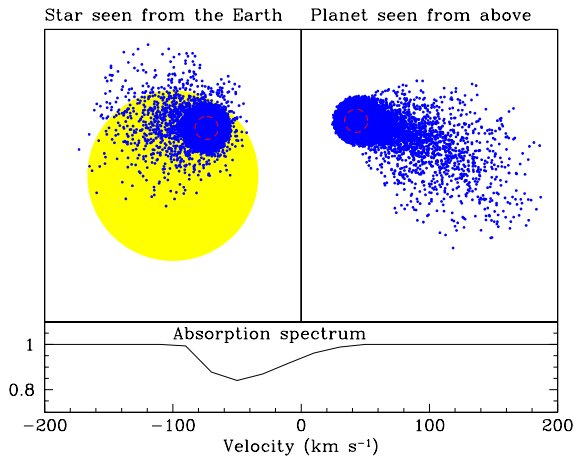}{5cm}
{-90}{30}{30}{-120}{170} 
\caption{A numerical simulation of
hydrogen atoms sensitive to radiation pressure 
(0.7 times the stellar gravitation) above an altitude
of 0.5 times the Roche radius where the density is assumed to be
2$\times$10$^{5}$~cm$^{-3}$ is presented here. It corresponds to
an escape flux of $\sim 10^{10}$~g~s$^{-1}$. The mean
ionization lifetime of escaping hydrogen atoms is 4 hours. The model
yields an atom population in a curved cometary like tail.}
\end{figure}

To evaluate the actual escape flux, one has thus to estimate the
vertical distribution of hydrogen atoms up to the Roche limit, in
a more realistic heated up and tidally extended upper atmosphere.

\subsection{The source of atomic hydrogen}

\hi\ is produced through H$_2$ photo--dissociation by extreme UV
(EUV) radiation at the top of the atmosphere and through
photolysis of hydrocarbons at the bottom. With an EUV heating
source of 160~erg~cm$^{-2}$.s$^{-1}$ on top of the atmosphere, in
principle one can derive the amount of \hi\ atoms at the base of
the upper atmosphere, in the region that could be considered as
the source of atoms. Using realistic condition for the bottom of
the atmosphere of HD209458b, Liang et al. (2003) have shown that
the atomic hydrogen could be as high as 10\%\ of H$_2$ while in
the Solar system giant planets it is of the order of 1\% .This is
one input parameter to be evaluated carefully.

\subsection{The upper atmospheric temperature}

Earlier estimates of the atmospheric escape flux concluded that
the mass loss was not significant. However these estimates did not
properly consider the temperature rise in the planetary upper
atmosphere as it is observed in all Solar system planets. 
For example, the radiative equilibrium,
effective temperatures \teff\ for the Earth and Jupiter are around
253~K and 150~K respectively while upper thermospheric and
exospheric temperatures \tup\ are $\sim$1000~K for both planets
(Chamberlain \&\ Hunten 1987).

Burrows \& Lunine (1995) estimated that the first ``hot Jupiter''
discovered, 51~Peg~b, was comfortably within its Roche tidal lobe
and thus that classical Jeans evaporation was insignificant.
However they recognized that ultraviolet heating may play a role by
mentioning that production of hot ions and atoms could be a more
promising escape mechanism. By scaling from the case of Jupiter
they evaluate that 10$^{34}$~atoms.s$^{-1}$ could possibly escape
from 51~Peg~b ($\sim$10$^{10}$~g.s$^{-1}$), a value similar to the
Vidal--Madjar et al. (2003) lower limit.

Using the atmospheric effective temperature \teff=1300~K as
the reference temperature, Guillot et al. (1996) concluded also
that ``hot Jupiters'' even at such ``high'' temperatures are much
too compact for classical Jeans escape. They also recognized that
escape of hot ions is a possible and more likely process.

First Chassefi\`ere et al. (1996) considered upper atmospheric
heating and evaluated that exospheric temperatures as high as
\tup$\sim$10$^4$~K were possible leading to upper atmospheres
extended up to 10~\rjup . Such possibilities were again mentioned
by Schneider et al. (1998) while Coustenis et al. (1998) underlined
again that upper atmospheric temperatures of extrasolar planets
should be quite different from the classical \teff\ evaluations.
Mentioning upper atmospheric heating by stellar EUV,
they evaluated \tup$\sim$4600~K in the case of 51~Peg~b.

In the case of HD209458b, Seager \& Sasselov (2000) made a 
detailed lower atmospheric model. They evaluated the lower
atmosphere effective temperature to be \teff=1350~K. They did not
consider any upper atmospheric heating. Following a parallel
approach, Brown (2001) found \teff=1400~K. He however clearly
stated that the model calculation did not involve He, H$_2$ and H
which can absorb UV photons and are very abundant, suggesting that
an upper atmospheric heating could be expected.

Finally more recently Lecavelier des Etangs et al. (2004) evaluated
the upper atmospheric temperature of HD209458b by taking into
account FUV and EUV heating and different cooling mechanisms. 
They evaluated \tup$\sim$11000~K.

Note however that the observed high upper atmospheric temperatures
in the Solar system are not all explained. In particular EUV solar
fluxes are able to heat up the upper atmospheres of the giant
planets by no more than few tens of degrees ($\sim$25~K in the
case of Jupiter). Clearly other mechanisms as heating by
solar wind particles
and/or by gravity waves within the upper atmospheres may
contribute (Hunten \&\ Dessler 1977, McConnell et al.
1982). This suggests that the Lecavelier des Etangs et al. (2004)
estimate could be a lower limit but knowing that more heating
produces more escape and thus more cooling, more heating could
have a small impact on the actual upper atmospheric temperature of
the planet (see also H\'ebrard et al. 2004).

\subsection{Turbulent transport}

The transport of atomic hydrogen at high altitudes could be done
via thermal diffusion but also through turbulent transport, 
as in the case of the Solar system giant planets. This
transport is commonly described by the classical eddy diffusion
coefficient parameter. It could vary by a huge 10000 factor, from
$10^4$~cm$^2$.s$^{-1}$ at Uranus to $10^6$~cm$^2$.s$^{-1}$ at
Jupiter to more than $10^7$~cm$^2$.s$^{-1}$ Saturn and finally
$10^8$~cm$^2$.s$^{-1}$ at Neptune (see e.g. Atreya, 1986). Consequently, 
much more atomic hydrogen could be present at the source level (the lower 
thermosphere), and could considerably
increase the amount of atomic hydrogen in the upper layers of the
atmosphere. Eddy diffusion could thus change the vertical
transport tremendously and possibly increase the amount of atomic
hydrogen escaping the planet. In the case of the HD209458b,
according to the probable extreme situation of that giant planet,
one could expect such transport mechanisms. By ignoring 
the unconstrained turbulent transport, we will thus obtain 
a lower limit of the actual escape flux from the planet.

\begin{figure}
\plotfiddle{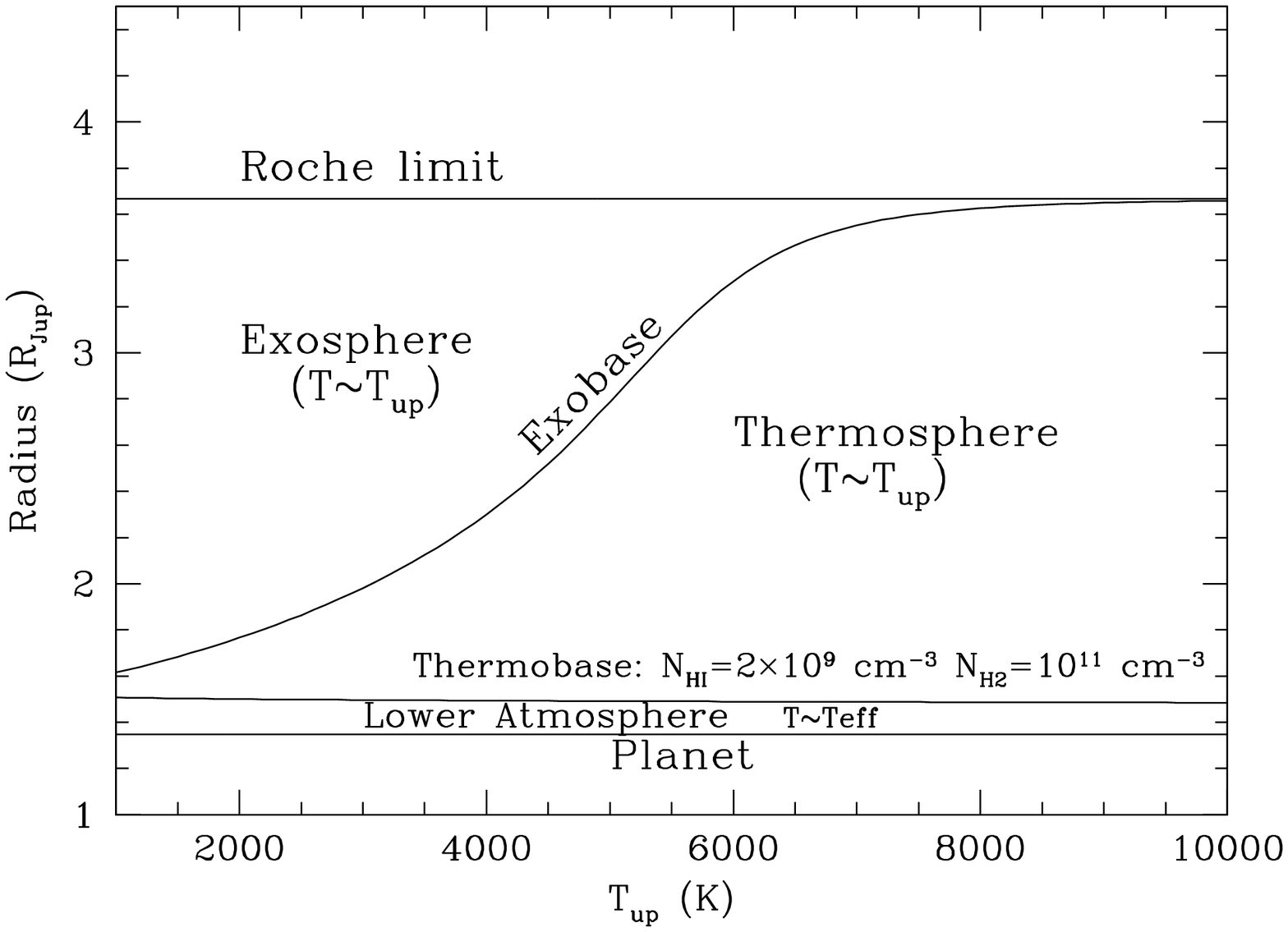}{0cm}{0}{31}{31}{-200}{-212}
\plotfiddle{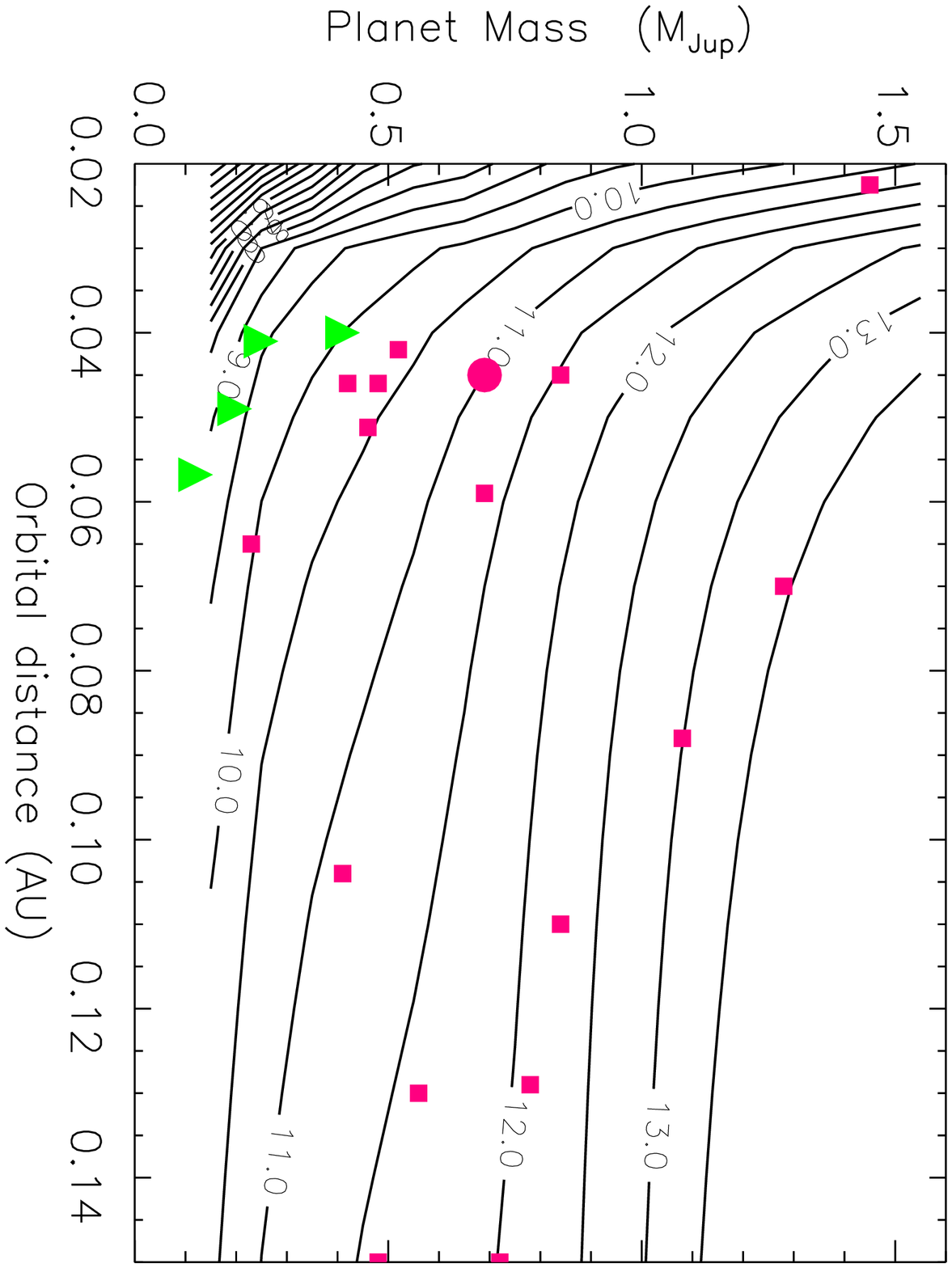}{4.0cm}{90}{28}{28}{200}{-19}
\caption{{\bf Left:} The upper atmosphere structure 
as a function of the exospheric temperature T$_{up}$. 
For exospheric temperatures above $\sim$7\,000~K
escape should be very efficient. {\bf Right:} Taken from Lecavelier
des Etangs et al. (2004). Contour plot of the decimal logarithm of
the planet life time
as a function of the mass and orbital distance.
The squares show the positions of detected planets. ``Osiris'' is
shown by a circle. The planets around HD49674, HD46375, HD76700,
and HD83443 are shown by triangles. We see that the life time of
these four planets are about 10$^9$years. These planets must have
lost a large fraction of their hydrogen and might be remnants of
former hot Jupiters. 
}
\end{figure}

\subsection{The atomic hydrogen escape flux}

Estimating the source of atomic hydrogen, the upper atmospheric
temperature, taking into
account tidal forces and ignoring turbulent transports, Lecavelier des
Etangs et al. (2004) were able to evaluate the vertical
distribution of atoms in the upper atmosphere (Fig.~4).
By extending the exobase concept to the level with no collisions 
up to the Roche limit (the exobase is usually defined by
the level above which particles have no more collisions),
they evaluated the number of atoms escaping the planet.
They found that for upper atmosphere temperatures above $\sim$7000\,K,
the escape flux is larger than the flux derived from Lyman-$\alpha$
observations.
But higher exospheric temperatures could be reached producing escape fluxes
possibly as high as $\sim$10$^{35}$~atoms.s$^{-1}$
($\sim$10$^{11}$~g.s$^{-1}$), i.e. about ten times larger than the
minimum value evaluated from the observations. Such a high
escape flux could put the escape into the blow-off regime in
which all atmospheric species are carried away with atomic and
molecular hydrogen (see e.g. Watson et al. 1981; Kasting \&\
Pollack 1983; Chamberlain \&\ Hunten 1987).

\section{Lifetime of an evaporating planet}

This raises the question of the lifetime of the evaporating extrasolar
planets which may be comparable to stellar lifetimes. If so, the
``hot Jupiters'' could evolve faster than their star, eventually
becoming ``hot hydrogen--poor
Neptune--mass planets''. The evaporation process, more efficient
for planets close to their star might also explain the very few
detections of ``hot Jupiters'' with orbiting periods shorter than
3~days (e.g., Konacki et al. 2003; Udry et al. 2003).

If the escape evaluation made by Lecavelier des Etangs et al.
(2004) is correct, the planetary lifetimes could be evaluated as a
function of both the star--planet distance and the planet initial
mass (Fig.~4). This shows that very close to their stars, 
planets may have short lifetime. They could evolve into
planets with hydrogen--poor atmospheres, or even with no atmosphere at all,
their inner core being directly exposed.

\section{Conclusion and naming planets}

The unique case of the transiting planet HD209458b allowed the
first exoplanets atmospheric studies. 
%
%
The case of HD209458b being so unique in the domain of exoplanets
studies and its current name so uneasy to use (and unappealing to
everyone outside the scientific community), it seems reasonable to
propose another name in addition to the present ones already given
by the standard nomenclature to its star (to be simply followed by
the letter ``b'' to name the planet). The star names are as listed
in the Centre de Donn\'ees Stellaires (CDS): V* V376 Peg, AG+18
2243, AGKR 19726, BD+18 4917, GSC 01688-01821, HD 209458, HIC
108859, HIP 108859, PPM 141002, SAO 107623, SKY\# 42008, TYC 1688-
1821-1, uvby98 100209458, and YZ 18 9012.

The idea is to use the Egyptian God name ``Osiris'' who was killed
and parts of his body spread over the whole Egypt. To have him
come back to life his sister Isis searched for all the pieces. She
found all of them but one and so, this God having lost part of his
body, could perfectly represent HD209458b, the evaporating planet.
This was proposed at the XIX$^{th}$ IAP Colloquium, Extrasolar
Planets, Today \& Tomorrow, on July 1, 2003.


\end{document}